\documentclass[amsmath,amssymb,physrev]{revtex4} \usepackage{url}
\usepackage{enumerate} % \usepackage{showkeys} \usepackage{times}
\usepackage{graphicx}% Include figure files \usepackage{bm}% bold math
\def\mode#1#2{#1^{(#2)}} 
\def\eqnn#1{Eq.~(\ref{eq:#1})}
\def\sect#1{Sect.~\ref{sec:#1}} 
\def\eqnb#1{(\ref{eq:#1})}
\def\eqnn#1{Eq.~\eqnb{#1}} 
\def\figno#1{Fig.~\ref{fig:#1}}
\def\tabno#1{Table~\ref{tab:#1}}

\def\Re{{\rm \,Re}\,} \def\Im{{\rm \,Im}\,}

\begin{document} %
\title{ Stable and unstable periodic orbits in the
one dimensional lattice $\phi^4$ theory } 
\author{Kenichiro Aoki\footnote{E--mail:~{\tt ken@phys-h.keio.ac.jp}.}} 
\affiliation{${}^*$Research and Education Center for Natural Sciences and Hiyoshi
Dept. of Physics, Keio University, Yokohama 223--8521, Japan\\ }
\begin{abstract} 
Periodic orbits for the classical $\phi^4$ theory on
the one dimensional lattice are systematically constructed by
extending the normal modes of the harmonic theory, for periodic, free
and fixed boundary conditions. Through the process, we investigate
which normal modes of the linear theory can or can not be extended to
the full non-linear theory and why. We then analyze the stability of
these orbits, clarifying the link between the stability, parametric
resonance and the Lyapunov spectra for these orbits.  The construction
of the periodic orbits and the stability analysis is applicable to
theories governed by Hamiltonians with quadratic inter-site potentials
and a general on-site potential. We also apply the analysis to
theories with on-site potentials that have qualitatively different
behavior from the $\phi^4$ theory, with some concrete examples.
\end{abstract}
\maketitle
\section{Introduction}
\label{sec:intro} 
Periodic motion is truly a classic topic in
classical mechanics, with the harmonic oscillator being a typical
example. Periodic motions are ubiquitous in nature, and physical
systems, such as a pendulum, often also contain anharmonicity at some
level. Therefore, linear and non-linear oscillations have been studied
for a long time\cite{Landau}. For systems with many degrees of
freedom, the harmonic theory is well understood and can be analyzed in
terms of normal modes of the theory, by using their linear
combinations.  However, when the system has many continuous degrees of
freedom and is anharmonic, there is still some progress to be made in
the systematic study of their periodic orbits.

In this work, we systematically construct periodic orbits in
non-linear lattice models by extending normal modes of the harmonic
theory. The class of models we study are conservative, Hamiltonian
systems, with quadratic inter-site potentials and general on-site
potentials. In these models, the origin of the non-linearity is
contained in the localized on-site potentials. In particular, we
investigate the $\phi^4$ theory, with periodic, fixed and free
boundary conditions, in some detail.
The main purpose of this work is to explicitly work out how the
various aspects of the non-linear dynamics come together in
generalizing the normal modes to these class of models.
Through the construction of the periodic solutions in the non-linear
theories, we clarify which modes in the linear theory can and can not
be extended.  We then analyze their stability from a dynamical systems
viewpoint.  In the process, we compute the Lyapunov spectra along
these periodic orbits, and show quantitatively how they are related to
the stable and unstable modes that appear, as the energy corresponding
to the periodic orbit is changed.  Furthermore, a general method of
finding extensions of normal modes in lattice theories with quadratic
inter-site and general on-site potentials is constructed, and
explained with examples.

While the dynamics of the theories we study here are of interest on
their own, these theories arise naturally as discretized versions of
the continuum field theory, of which $\phi^4$ theory is a typical
case. $\phi^4$ theory has been studied from various points of view on
the lattice: Classically, the transport properties of the theory have
been investigated from statistical mechanics viewpoint at finite
temperatures\cite{AK1,Hu} and their relation to dynamical systems
aspects of the theory, including the Lyapunov spectrum and dimensional
loss\cite{dloss}, with thermostats.  
The on-site potentials of the models we study destroy the shift
symmetry properties of the fields (cf. \sect{phi4}) that exist in well
studied models such as the FPU model\cite{FPU}, leading to qualitatively
different dynamical behavior, such as the bulk behavior of transport
coefficients\cite{ThermoRev}.
The chaotic properties and the
Lyapunov exponents of the non-thermostatted $\phi^4$ theory, as well
as some of its periodic orbits have also been
investigated\cite{HA1}. The $\phi^4$ theory, including quantum
effects, has been studied in such topics as
triviality\cite{triviality}, and non-perturbative aspects of particle
physics phenomenology\cite{phenomenology}, since the $\phi^4$ theory
is a part of the Standard Model. While the physics of the $\phi^4$
theory investigated in this work is classical, understanding of the
classical theory is also important to the understanding of the quantum
theory, and furthermore, classical solutions can be an important
contributing factor in quantum theories.

The periodic orbits we study are so-called ``non-linear normal modes''
of the class of non-linear models.  Non-linear normal modes have been
studied extensively for some time and various general properties have
been
established\cite{Rosenberg,NLNM0,SanduskyPage,Flach,NLNMRev,ChechinRev},
and investigated in models such as the FPU
model\cite{BudinskyBountis,stabFPU,NLNMFPU}.  Most of the explicit
work conducted so far seems to focus on theories with non-linear
inter-site couplings. These properties can, for instance, represent
non-linearity in the elastic response of materials and can be of
practical importance. Accordingly, much applied research has been
performed on the subject\cite{NLNM1}. We believe that our work, which
concentrates on models with quadratic inter-site potentials, with
non-linear couplings that are local, contributes results complementary
to the current research in the dynamics of non-linear systems with
many coupled degrees of freedom.
\section{Lattice $\phi^4$ theory in one spatial dimension}
\label{sec:phi4}
Let us review the $\phi^4$ theory on an one-dimensional lattice with
$N$ sites, in brief, partly to fix the notation. The Hamiltonian of
the theory is
\begin{equation}
    \label{eq:ham} H=\sum_{j=1}^N  {p_j^2\over2}
      +\sum_{j=1}^{N
      -1}{(q_{j+1}-q_j)^2\over2}+H_{\rm B}+\sum_{j=1}^N{q_j^4\over4}
    \quad,
\end{equation}
where the potential terms at the ends, $H_{\rm B}$, depend on the boundary
conditions as
\begin{equation}
    \label{eq:lb} 
    H_{\rm B}= {1\over2}(q_{N}-q_1)^2\quad\hbox{(periodic bc)},
    \quad H_{\rm B}= {1\over2}\left(q_{N}^2+q_1^2\right)
    \quad\hbox{(fixed bc)},\quad H_{\rm B}= 0\quad\hbox{(free bc)}.
\end{equation}
The non-linearity of the system is provided by the quartic on-site
potential, or the tethering potential. 
The on-site potential destroys the shift symmetry (shifting all $q_j$
by a constant), which exist in other well studied non-linear models
such as the FPU model, leading to a different dynamical behavior for them.
The system has a quartic
coupling of essentially one, and is not a weakly coupled theory, in
general.  The equations of motion for the theory are accordingly,
\begin{equation}
    \label{eq:eom}
    \dot q_j =p_j,\quad     \dot p_j= q_{j+1}+q_{j-1}-2q_j -q_j^3
    \quad j= 1,2,\cdots,N\quad.
\end{equation}
The boundary conditions may be specified as,
\begin{equation}
    \label{eq:eomBC}
    q_0=q_N,q_{N+1}=q_1\ \hbox{(periodic)},\quad
    q_0=q_{N+1}=0\ \hbox{(fixed)},\quad
    q_0=q_1, q_{N+1}=q_N\ \hbox{(free)}\quad.
\end{equation}
We note the fixed boundary condition  by itself also breaks the shift symmetry.
The two boundaries may also have different conditions, which shall not
be considered here.
\section{Systematic construction of periodic orbits in $\phi^4$ theory}
\label{sec:construction}
In this section, we construct a  class of periodic solutions in
the $\phi^4$ theory on the lattice, based on the normal modes of the
harmonic theory.  
A more general study of periodic solutions using powerful group theoretical
methods have been conducted\cite{ChechinRev,bush0,bush1}, and have been
applied to models such as the FPU model\cite{NLNMFPU}. Here, we
briefly explain a more elementary approach to the solutions which
hopefully provides some different insight, and derive the results
needed later. The class of solutions we study are sometimes called
non-linear normal modes or one dimensional
bushes\cite{bush0,bush1,ChechinRev}.

Suppose that the $N$ linearly independent normal
modes of the harmonic chain are,
\begin{equation}
    \label{eq:normalmode}
    y_j=\mode {a_j}m \cos \omega^{(m)} t, \quad
    m=0,1,\cdots,N-1.
\end{equation}
These  normal modes satisfy the {\em linear} equations of motion, 
\begin{equation}
    \label{eq:nmeom}
    \ddot y_j=  y_{j+1}+y_{j-1}-2y_j =\left(    a_{j-1}^{(m)}+
      a_{j+1}^{(m)}- 2a_j^{(m)}\right) \cos \omega^{(m)} t 
    = -(\omega^{(m)})^2 a_j^{(m)} \cos \omega^{(m)} t \quad.
\end{equation}
The solutions to these linear equations of motion may be found using
the coefficients of the form, $\mode{a_j}m=\Re({\rm const.}\times
\exp(i\mode kmj))$, and the harmonic frequencies can be found as,
\begin{equation}
    \label{eq:omegaSin}
    \mode \omega m=2\left|\sin {\mode km\over2}\right|\quad,
\end{equation}
% This procedure works for all the boundary conditions we consider, and
The values of $\mode km$ depend on the boundary conditions.

We shall analyze which of these solutions extend to the $\phi^4$
theory, for {\it general} $N$. Specifically, we look for solutions of
the non-linear theory, in which all the coordinates undergo the same
motion, except possibly the amplitudes of the motion.  Using the
ansatz,
\begin{equation}
    \label{eq:normalmode44}
    q_j=a_j^{(m)} f^{(m)}(t)\quad,
\end{equation}
the equations of motion reduce to
\begin{equation}
    \label{eq:normalmode4}
    a_j^{(m)}\ddot  f^{(m)}(t) =     -(\omega^{(m)})^2 a_j^{(m)}
    f^{(m)}(t) -    ( a_j^{(m)})^3( f^{(m)}(t))^3\quad.
\end{equation}
These equations are consistent if and only if the non-trivial
equations are independent of $j$.  This is satisfied when the square
of non-zero coefficients, $a_j^{(m)}$, are independent of $j$,
\begin{equation}
    \label{eq:cond}
     (a_j^{(m)})^2  = C\quad \hbox {when}\quad a_j^{(m)}\not=0\quad,
\end{equation}
in which case, the equations of motion reduce to an ordinary
differential equation for a non-linear oscillator, $z(t) = \mode{a_j}m
\mode f m(t)$
\begin{equation}
    \label{eq:eomf}
    \ddot  z =     -(\omega^{(m)})^2    z -  z^3\quad.
\end{equation}
The solutions to these equations are periodic in time.  
% This provides us with a simple criterion for whether the normal modes of the
% harmonic theory can be extended to periodic orbits in the non-linear
% $\phi^4$ theory and provides us with the non-linear differential
% equation for the orbit, when this is possible.  
Though seemingly simple, the procedure has reduced the non-linear
coupled $2N$ first order differential equations to just one second
order differential equation.  While there seem to be a few definitions
of the ``non-linear normal modes'', periodic orbits constructed above
are non-linear normal modes in the strict sense\cite{Rosenberg}. The
motion of the coordinates are ``similar'' and synchronous --- they all
undergo identical motion with respect to time, except for their
amplitudes.  It is important to note that the periodic orbits we have
found here exhausts all the synchronous oscillations that can have
arbitrary overall amplitudes. This is because all such modes should
still be synchronous when the overall amplitude goes to zero.  In this
limit, the dynamics become harmonic, and the motions need to reduce to
the standard linear normal modes.  Obviously, changing the overall
amplitude of a motion is equivalent to changing its energy.  The
non-linearity of the theory manifests itself in the motions
themselves; the shapes and the periods of the orbits depend on their
energy, which is a qualitatively different behavior from the linear
theory.

Clearly, this kind of construction is valid for any equations of
motion, in which the couplings between the sites lead to linear terms
in the equations of motion, and the only non-linearities are due to
the on-site potential. In particular, this construction works for any
boundary condition, since the boundary conditions change only the
linear parts of the equations of motion.  Another point evident in the
above derivation is that this construction can be generalized to a
theory with any on-site potential. 
We shall investigate the dynamics
of on-site potentials other than the $\phi^4$ theory in
\sect{extension}.

\begin{figure}[htbp]
    \centering
    \includegraphics[width=8.6cm,clip=true]{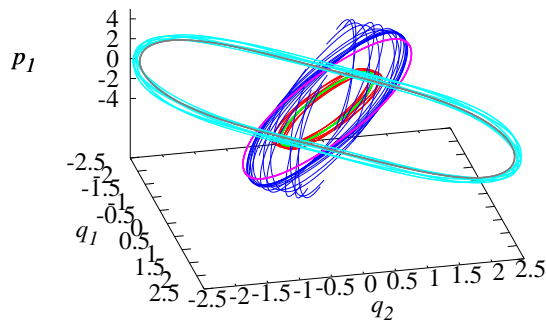}    \\
    \caption{Periodic orbits and their perturbations for $N=2$: The
      symmetric trajectory for $E/N=8$ (brown), antisymmetric
      trajectories for $E/N=2$(green) ,8(magenta) and their
      perturbations (cyan, red, blue, respectively). Antisymmetric
      trajectory for $E/N=8$ is dynamically unstable, while others are
      not, which is visible.  All trajectories, $(q_1,q_2,p_1)$, were
      started with $q_1=q_2=0$ and were followed for the same amount
      of time, $\Delta t=40$. Perturbed orbits were obtained by
      increasing the initial $p_1$ value by 10\,\%. }
    \label{fig:N2Traj}
\end{figure}
Let us consider some simple examples :
\paragraph{Symmetric orbit for periodic boundary conditions, for
  any $N$}
In the harmonic theory, when the boundary condition is periodic, a
trivial solution with an arbitrary shift by a constant is a solution
to the equations of motion, with $\omega^2=0$. This satisfies the
condition, \eqnn{cond}, so that this solution $q_e$, which we shall
call ``symmetric'' can be extended to the non-linear $\phi^4$ theory,
with the equations of motion\cite{HA1}
\begin{equation}
    \label{eq:sym}
 \ddot q_e=-q_e^3\quad.
\end{equation}
While the original solution of the harmonic theory was just a constant
shift, it should be noted that this equation is a non-linear equation
and hardly trivial.  
This contrasts with other models with non-linearities only in the
inter-site couplings, such as the FPU model. In such models, these
solutions are trivial.
\paragraph{Antisymmetric orbits for periodic boundary condition,
  when $N$ is even}
Without the on-site potential, the ``antisymmetric'' normal mode
$q_{2j}=-q_{2j+1}=q_o,p_{2j}=-p_{2j+1}=p_o$ (any $j$) exists. This
satisfies the condition \eqnn{cond} so that the equations of motion
can also be extended to the $\phi^4$ theory as\cite{HA1}
\begin{equation}
    \label{eq:antiSym}
    \ddot q_o=-q_o^3-4q_o\quad.
\end{equation}

While these constructions might seem simple, not all normal modes can
be extended to the non-linear theory and the resulting equations are
non-trivial. This distinction between the linear and non-linear
theories is quite clear, for instance, when we consider the $N=3$
system with periodic boundary conditions.  In this case, the three
linearly independent normal modes in the harmonic theory have
amplitudes of constant times $(1,1,1),(1,-1,0),(0,1,-1)$, for the
three coordinates. All these modes can be extended to the $\phi^4$
theory. However, a mode that can be obtained from two of the modes
which have the same frequency, with amplitudes, $(2,-1,-1)$, is also a
normal mode in the harmonic theory, that can {\it not} be extended to
the non-linear theory, since it does not satisfy the condition,
\eqnn{cond}. On the other hand, a solution that can be obtained as a
sum of the amplitudes that satisfy the condition, \eqnn{cond}, is not
a simple sum of the solutions and is another non-trivially different
solution in the $\phi^4$ theory, since the equations of motion are
non-linear.  Some symmetric and antisymmetric orbits, along with their
perturbed trajectories are shown for the $N=2$ lattice with periodic
boundary conditions in \figno{N2Traj}.

While we mentioned some simple examples above, we list the normal
modes that can be extended for the three boundary conditions,
periodic, fixed and free, up to $N=9$ in \tabno{modesBC}.  The
results depend on the boundary conditions in an interesting manner, as
we explain.
The general theory of these modes have been established
previously\cite{bush0,bush1,ChechinRev}, using group theoretical methods.
The results in \tabno{modesBC} were obtained by solving the condition,
\eqnn{cond}. The general case, including these, is explained
below. The symmetric orbits explained above, are not shown in this
table, and exists for any $N$ when the boundary conditions are periodic or
free.  The modes in \tabno{modesBC} are listed simply with the
amplitudes of each $q_j$ and this can be multiplied by any common
constant value, and still be a periodic orbit. We have only listed
modes which are inequivalent. In particular, the modes which are
equivalent by just by shifting the oscillator in the periodic case, or
by reflection (oscillator $j\leftrightarrow N-j$) are not listed in
the tables.
\begin{table}[htbp]
    \centering
    \begin{tabular}[t]{r|c|c}\hline
        $N$&$\omega^2$& $a_j$ (periodic bc)\\\hline
        2&4&$(1,-1)$\\\hline
        3&3&$(1,-1,0)$\\\hline
          4&2& $(1,1,-1,-1)$\\
          &2& $(0,1,0,-1)$\\
        &   4& $(1,-1,1,-1)$\\\hline
        6&1& $(0,1,1,0,-1,-1)$\\
        &3& $(1,-1,0,1,-1,0)$\\
        &4& $(1,-1,1,-1,1,-1)$\\\hline
        8&2& $(1,1,-1,-1,1,1,-1,-1)$\\
        &2& $(0,1,0,-1,0,1,0,-1)$\\
        &4& $(1,-1,1,-1,1,-1,1,-1)$\\\hline
        9&3& $(1,-1,0,1,-1,0,1,-1,0)$\\\hline
    \end{tabular}\quad
    \begin{tabular}[t]{r|c|c}\hline
        $N$&$\omega^2$& $a_j$ (fixed bc)\\\hline
        2&1&$(1,1)$\\
        &3&$(1,-1)$\\\hline
        3&2&$(1,0,-1)$\\\hline
          5&1& $(1,1,0,-1,-1)$\\
          &2& $(1,0,-1,0,1)$\\
          &3& $(1,-1,0,1,-1)$\\\hline
        7&   2& $(1,0,-1,0,1,0,-1)$\\\hline
        8&1&$(1,1,0,-1,-1,0,1,1)$\\
        &   3& $(1,-1,0,1,-1,0,1,-1)$\\\hline
        9&2& $(1,0,-1,0,1,0,-1,0,1)$\\\hline
    \end{tabular}\quad
    \begin{tabular}[t]{r|c|c}\hline
        $N$&$\omega^2$& $a_j$ (free bc)\\\hline
        2&2&$(1,-1)$\\\hline
        3&1&$(1,0,-1)$\\\hline
          4&2& $(1,-1,-1,1)$\\\hline
        6&1& $(1,0,-1,-1,0,1)$\\
        &2& $(1,-1,-1,1,1,-1)$\\\hline
        8&2&$(1,-1,-1,1,1,-1,-1,1)$\\\hline
        9&1& $(1,0,-1,-1,0,1,1,0,-1)$\\\hline
    \end{tabular}
    \caption{Rescaled amplitudes for the non-linear periodic modes up
      to $N=9$ for periodic (left), fixed (middle) and      free
      (right) boundary conditions. In addition, the symmetric solutions
      $a_j=1$ (any $j$) with $\omega^2=0$ exist for all $N$ in the periodic
      and free boundary cases, which are not shown. }
    \label{tab:modesBC}
\end{table}

\begin{table}[htbp]
    \centering
    \begin{tabular}[t]{l|l|c|c|c}\hline
        label&$\omega^2$&period & amplitudes & mode\\\hline
        0&  0&   1& $(1)$&\includegraphics[height=8mm,clip=true]{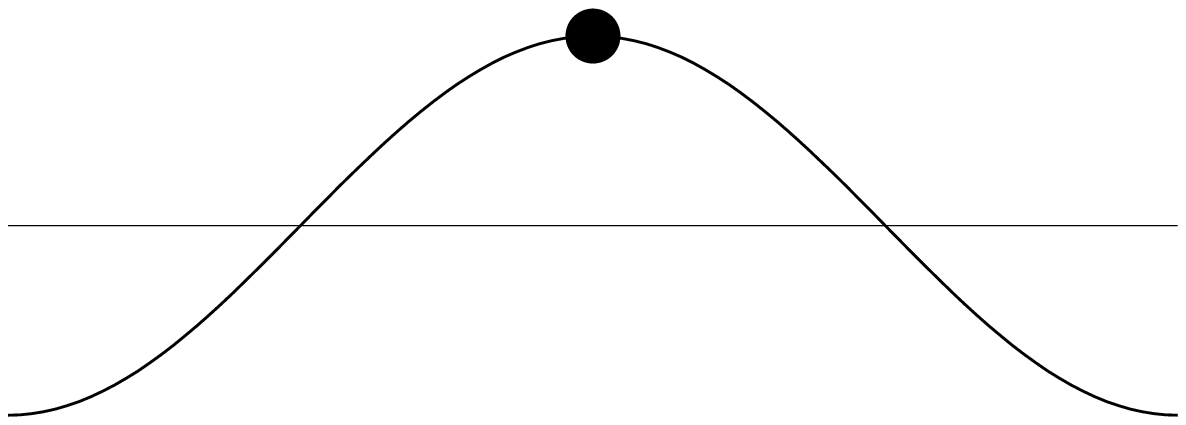}\\
        1& 1&  6& $(0,1,1,0,-1,-1)$&\includegraphics[height=8mm,clip=true]{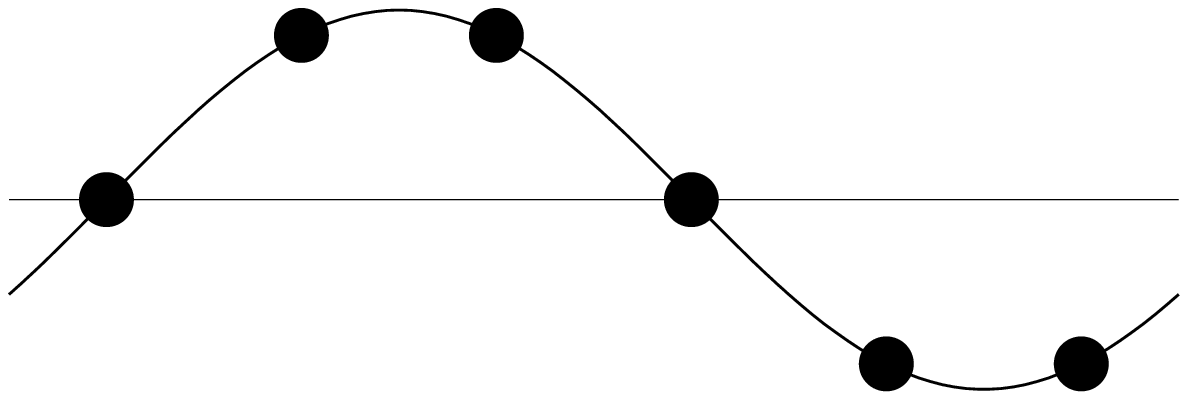}\\
        2a&2&   4& $(1,1,-1,-1)$&\includegraphics[height=8mm,clip=true]{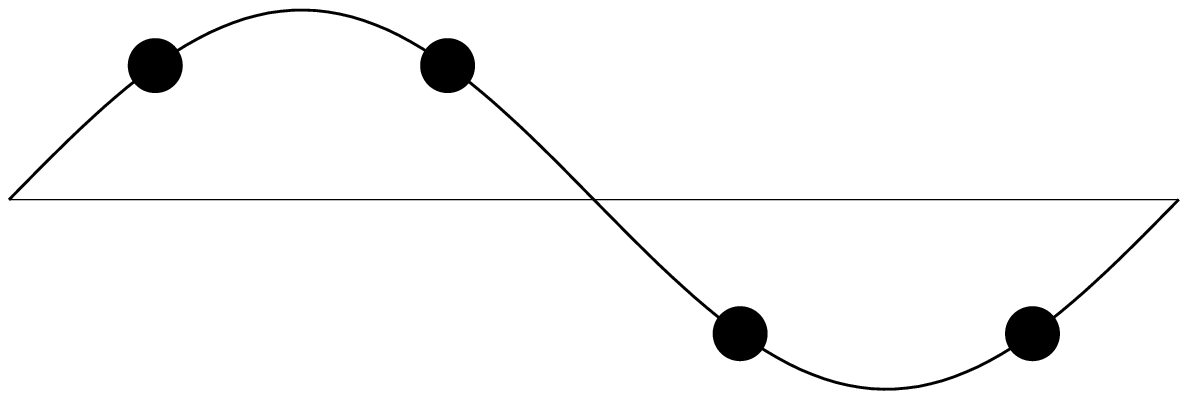}\\
        2b&2&   4& $(0,1,0,-1)$&\includegraphics[height=8mm,clip=true]{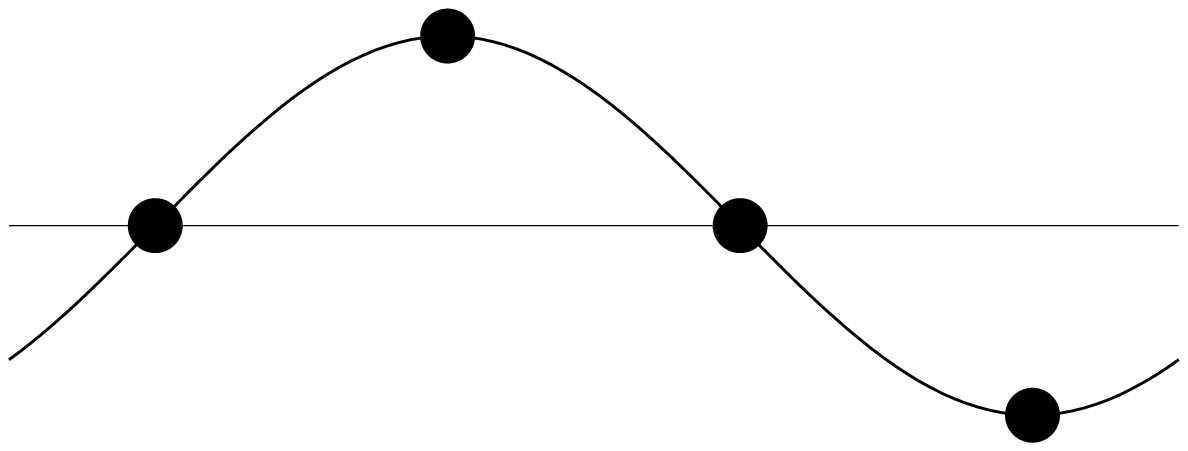}\\
        3&3&   3& $(1,-1,0)$&\includegraphics[height=8mm,clip=true]{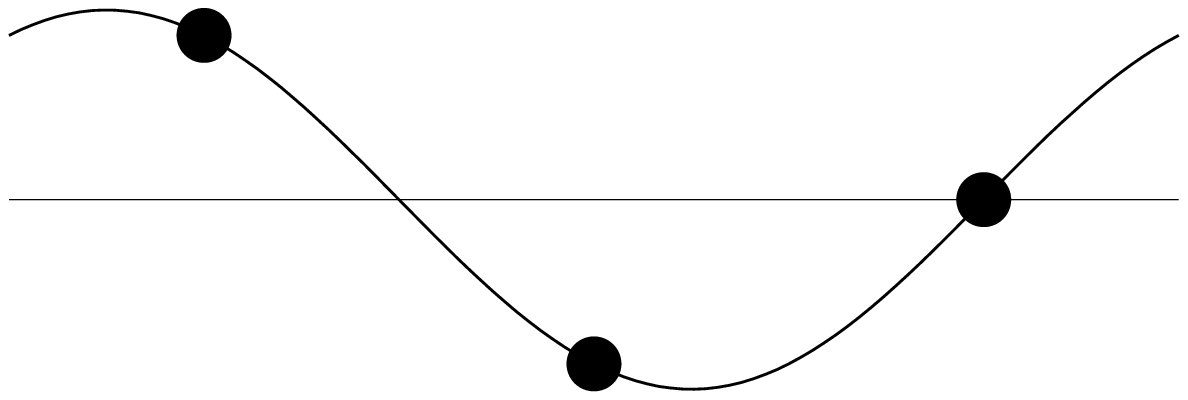}\\
        4&4&   2& $(1, -1)$ &\includegraphics[height=8mm,clip=true]{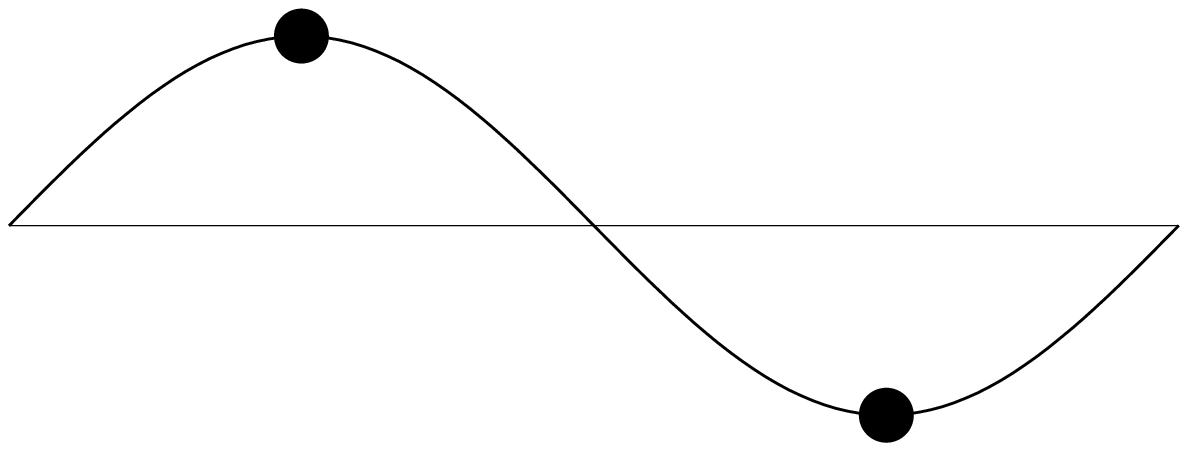}\\
        \hline
    \end{tabular}
    \caption{Allowed basic modes and their rescaled amplitudes. The
      modes  are labeled for reference
      (see text).}
    \label{tab:modes}
\end{table}
The general solution can be understood as follows: Since the
coefficients for the linear equations, \eqnn{nmeom}, are integers,
$\left(\mode \omega m\right)^2$ are also integers, under the condition
\eqnn{cond}. Due to the value of these coefficients, only $0,1,2,3$
and $4$ are possible, corresponding to the periods on the lattice of,
1,6,4,3 and 2, respectively, as tabulated in \tabno{modes}. To make
the relationship between $\omega^2$, periodicity and the mode clear, a
graphical representation of the basic modes are also shown.  %
All the modes are repetitions of these modes, except close to the
boundaries, as explained below.
Corresponding modes have been derived for the FPU model with periodic
boundary conditions\cite{NLNMFPU}.  While the results are similar, it
is interesting to note the the differences.  The $\phi^4$ theory,
unlike the FPU model, contains an on-site potential that leads to 
{\em non-linear local} interactions. This enforces the non-zero
amplitudes of the oscillations to be equal, as seen in \eqnn{cond},
which is not the case for  the FPU model.

Periodic boundary conditions are the simplest to understand. All the
solutions are repetitions of the modes in \tabno{modes}, and the
condition for the non-linear periodic solutions to exist is that $N$
is a multiple of a period in \tabno{modes}. $\omega^2$ will then have
the corresponding value. All the basic modes listed in \tabno{modes}
are allowed. The solutions up to $N=9$ are listed in
\tabno{modesBC}(left) can be all understood from this logic, and in
particular, only the symmetric solution exists for prime numbers
$N(>3)$.  The symmetric mode, where all the amplitudes are the same
value, exists for any $N$.

For fixed boundary conditions, the values at the boundaries need to be
$0$ as in \eqnn{eomBC}, so that $0$ needs to be contained in the
amplitudes and  only the modes 1,2b,3 are allowed (\tabno{modes}). The
periodicity of the zeros for these modes are 3,2,3, respectively, so
that the condition for these modes to exists is that $N+1$ is a
multiple of $2,3$ and the solutions are repetitions of the basic modes
1,2b,3,  adjusted so that the zeros are at the boundaries. In
some cases, only half of a basic mode might appear at the boundary, as
in the case the first mode for $N=8$ and the mode for $N=9$ in
\tabno{modesBC}(middle). A general non-linear mode can be constructed
this way, which are listed in \tabno{modesBC}(middle), up to $N=9$.
While the mode $(1,1)$ appears for $N=2$, unlike the symmetric modes
for the periodic and the free boundary conditions, it is a part of
mode 1 in \tabno{modes} with $\omega^2=1$. For other values of $N$, no
mode with all amplitudes being equal appears for fixed boundary
conditions.

For free boundary conditions, the two consecutive sites with the same
values need to appear, so that only the basic modes 0,1,2a in
\tabno{modes} are allowed. The symmetric mode (mode 0), which
obviously has this property, exists for any $N$.  The periodicity
between the identical consecutive values for modes 1,2a are 2,3,
respectively, so that the non-linear modes exist for $N$ being
multiples of 2,3. All modes can be understood in this manner and they
are listed up to $N=9$ in \tabno{modesBC}(right).
\section{Stability and instability of the periodic orbits}
\label{sec:stability}
The periodic orbits explained in the previous section have only two
first order degrees of freedom in essence, and might not seem
``chaotic''.  However, they can be unstable from a dynamical systems
perspective. In this work, we investigate the stability, or
lack thereof, of the periodic orbits, from this point of view.  When
the orbit is unstable, perturbations grow exponentially large with
time. Therefore, this instability appears as the positive maximal
Lyapunov exponent along the orbit. Obviously, the instability can also
be seen explicitly by following perturbed trajectories. If the
exponent is positive, a small perturbation on the orbit will cause the
trajectory to diverge exponentially from the periodic orbit.  While,
this tells us how to discriminate when the orbit is unstable, it does
{\it not} tell us {\it why} the orbit is unstable or stable. For this,
we now turn to an analysis of the perturbations around the orbit.
General theory of stability analysis has been studied previously and
have been performed explicitly for periodic orbits in the FPU
model\cite{stabFPU,bush1,ChechinRev}.

Partly to avoid confusion, it should be mentioned that even when the
orbits are dynamically unstable, some orbits can be stable from a
computational standpoint\cite{HA1}. This is a technically interesting
issue, perhaps of practical import, which we briefly explain: Any
numerical computation contains round-off errors and has only a finite
precision.  Therefore, one might expect that following unstable orbits
numerically for a long time is impossible, since any deviation will
force the trajectory to diverge exponentially with time from its
``true'' trajectory. However, somewhat surprisingly, some periodic
orbits can be followed for an arbitrarily long time. The reason for
this is that their symmetry properties are preserved to the last bit
in the data, with the appropriate coding and the use of compilers.
For instance, in the integration of the the symmetric and
antisymmetric orbits explained in previous section, the properties
$q_{2j}=\pm q_{2j+1}$ are fully preserved in the numerical integration.  While
it is unclear if this situation applies to all periodic orbits, it
applies also to other orbits we investigate below.  This property
allows us to follow periodic trajectories and compute Lyapunov spectra
averaged along them with precision, given enough computational
time. In this work, we used the fourth order Runge-Kutta routine for
integration and the method explained in \cite{Lyap2} to compute the
Lyapunov spectra.

Let us briefly summarize the properties of Lyapunov exponents, which
will be of use to us\cite{Lyapunov,HH1}. Lyapunov spectra have been
computed in various Hamiltonian
systems\cite{Lyap0,Lyap1,Lyap2,Lyap3,Lyap4,Lyap5,HH1}, including the
$\phi^4$ theory both thermostatted and not
thermostatted\cite{AK1,HA1}.  When one follows a trajectory in phase
space, the neighboring trajectories can diverge from (or converge to)
the original trajectory exponentially and their exponents per unit
time are called Lyapunov exponents. If we consider all the possible
different directions of the neighboring trajectories and average along
the original trajectory, their rates of divergence or convergence from
it, we obtain the Lyapunov spectrum.  The systems considered in this
work are all Hamiltonian systems, with the Hamiltonian having no
explicit time dependence. For $N$ pairs of coordinates and momenta,
$q_j,p_j\ (j=1,2,\cdots N)$, there are $2N$ Lyapunov exponents. The
spectrum of exponents is made up of pairs of the form $\pm \lambda$
($\lambda$: Lyapunov exponent) and is invariant under changing the
sign of all the exponents, due to the time reversal symmetry of the
system. Furthermore, at least one pair of exponents is zero, since the
trajectories are on a $2N-1$ dimensional constant energy surface. It
should be noted, that in this work, we follow periodic orbits, which
are localized in the phase space, and compute the Lyapunov spectra
along them. So the Lyapunov spectra obtained here are different from
the spectra obtained by averaging over the chaotic ``sea'' in the
phase space\cite{Lyap0,Lyap1,Lyap2,Lyap3,Lyap4,Lyap5,HH1}.
\subsection{$N=2$ system with  periodic boundary conditions}
\label{sec:N2P}
The simplest case to analyze the stability of the periodic orbits is
the $N=2$ system, since the $N=1$ system will only have zero Lyapunov
exponents, and hence no instability, from the properties referred to
above.  Below, this system with periodic boundary conditions is
analyzed in some detail. For $N=2$, there are only the ``symmetric''
orbit, in which both coordinates are the same, and the
``antisymmetric'' orbit, in which the coordinates are $(-1)$ times
each other.  These cases are instructive and enables us to clearly see
the mechanism behind the instability of the orbit, or lack thereof,
enabling us to extend this understanding to more general cases.  The
Lyapunov exponents for the periodic orbits has been computed in some
cases\cite{HA1}. It was found that the symmetric solutions,
\eqnn{sym}, seemed to have zero maximum Lyapunov exponent for any
energy, though numerical computations can not rule out small non-zero
exponents.
In contrast, the antisymmetric orbits, \eqnn{antiSym}, were found to
be stable at low energies, becoming unstable at higher energies.  This
stark contrast is intriguing, but its underlying physics was
unclear. We will see how this originates below.

The symmetries of the system are more conveniently viewed using the
coordinates 
\begin{equation}
    \label{eq:coord2}
    \chi={1\over2}\left(q_1-q_2\right),\quad
    \eta={1\over2}\left(q_1+q_2\right)\quad.
\end{equation}
Then, the equations of motion for the system can be reorganized into a
more convenient form as 
\begin{equation}
    \label{eq:eom2}
    \ddot \chi=-\chi\left(\chi^2+3\eta^2\right)-4\chi,
    \quad
    \ddot \eta=-\eta\left(\eta^2+3\chi^2\right)\quad.
\end{equation}
To analyze the problem of stability, we perturb around a general
classical solution, $\chi_0,\eta_0$, as
$\chi=\chi_0+\chi_1,\eta=\eta_0+\eta_1$. Keeping only the leading order
terms, we arrive at the equations for the fluctuations around the
solution, 
\begin{equation}
    \label{eq:fluct}
    \ddot  \chi_1=-3\left(\chi_0^2+\eta_0^2\right)\chi_1-4\chi_1-6\chi_0\eta_0\chi_1,\qquad
    \ddot \eta_1=-3\left(\eta_0^2+\chi_0^2\right)\eta_1-6\chi_0\eta_0\eta_1    \quad.
\end{equation}
\subsubsection{Symmetric mode fluctuations}
\begin{figure}[htbp]\centering
    \includegraphics[width=8.6cm,clip=true]{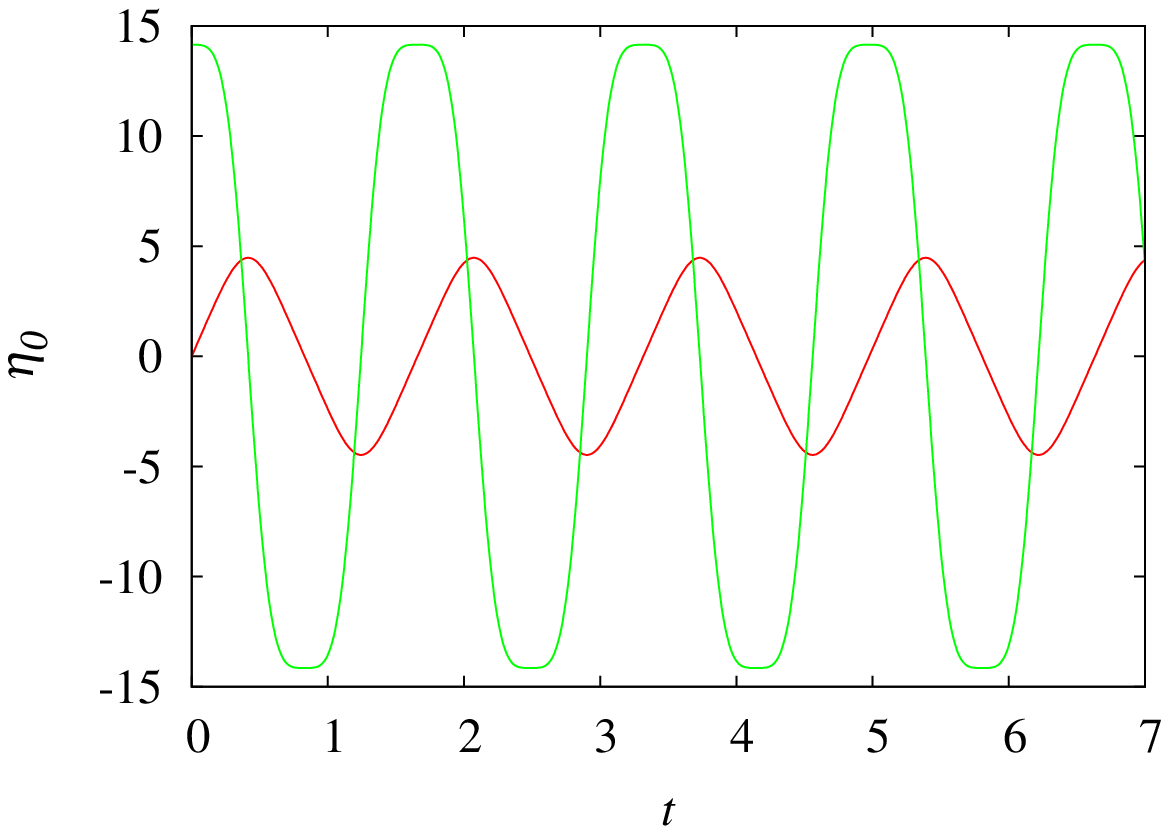}    \\
    \includegraphics[width=8.6cm,clip=true]{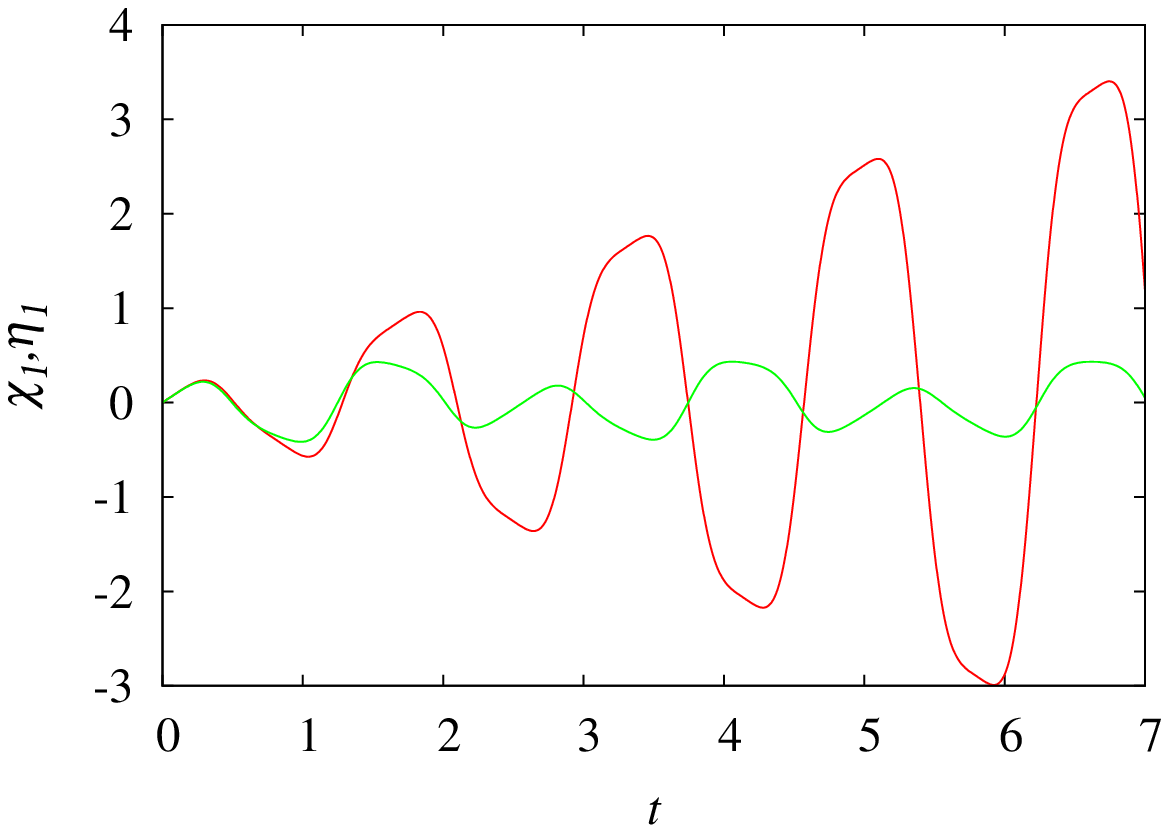}    
    \caption{$N=2$ symmetric orbit trajectories of $\eta_0$(red) and
      its momentum, $\dot \eta_0$(green), for $E/N=100$ (top
      figure). Time dependence of the perturbations $\eta_1$(red),
      $\chi_1$(green) around the symmetric mode, $\eta_0$ (bottom
      figure). While $E/N$ is relatively large, perturbation $\eta_1$
      grows only linearly with time, not exponentially, and $\chi_1$
      amplitude does not become larger. }
\label{fig:sym2}
\end{figure}
Let us first discuss small deviations from the symmetric orbit, 
$\chi_0=0$ and $\eta_0$ satisfying the non-linear oscillator equation,
\eqnn{sym}. The deviations from the orbit satisfy
\begin{equation}
    \label{eq:symFluct}
    \ddot  \chi_1=-\left(3\eta_0^2+4\right)\chi_1,\qquad
    \ddot \eta_1=-3\eta_0^2\eta_1\quad.
\end{equation}
These equations are those of harmonic oscillators with oscillation
frequencies that depend on time. The frequencies are clearly real for
both equations, so that there is no trivial exponential growth.  Yet
$\eta_0$ is periodic, so that solutions to these equations can exhibit
parametric resonance\cite{Landau}, which we now investigate.

While the equations \eqnn{symFluct} can be analyzed numerically, and
shall be done so below, it is worthwhile to study the mechanism
analytically.  Parametric resonance arises when the frequency of the
oscillation changes at a rate close to twice the base frequency, to
leading order.  The condition for such an instability for an oscillator
satisfying
\begin{equation}
    \label{eq:hoLandau}
    \ddot x= -\omega^2(t)x= -\omega_0^2\left(1+h\cos\gamma t\right)x\quad,
\end{equation}
is
\begin{equation}
    \label{eq:condLandau}
    \left|{\gamma\over2\omega_0}-1\right|<{h\over4}\quad.
\end{equation}

\def\etaMax{\eta_{\rm max}} While the basic motion, $\eta_0$ is not
sinusoidal, let us approximate $\eta_0$ by
$\sqrt{2}E_1^{1/4}\sin\Omega t$ to gain insight, where $E_1=E/N$ is
the energy per oscillator.  The frequency can be computed to be
$\Omega=(2\pi)^{3/2}E_1^{1/4}/\Gamma(1/4)^2$. One can then show
analytically that neither $\omega^2(t)=3\eta_0^2$ nor $
\omega^2(t)=3\eta_0^2+4$ satisfies the condition, \eqnn{condLandau},
for any $E_1$. In particular, we note that in the first case,
$\Omega/\omega_0$ is independent of $E_1$ and just a numerical
constant.  We can integrate the perturbations, \eqnn{fluct},
numerically and we find that the symmetric mode does not become
unstable regardless of the energy, which is consistent with what was
found analytically.  An example of the perturbations is illustrated in
\figno{sym2}. The initial conditions for the perturbations shown in
the plot are chosen to be
$(\xi_1(0),\dot\xi_1(0)),(\eta_1(0),\dot\eta_1(0))=(0,1)$ and the same
conditions will be used below for all perturbation mode analyses. The
equations for the perturbations are linear with respect to
$\xi_1,\eta_1$ so that rescaling these conditions just rescales the
solutions. 
\subsubsection{Antisymmetric mode fluctuations}
\begin{figure}[htbp]\centering
    \includegraphics[width=8.6cm,clip=true]{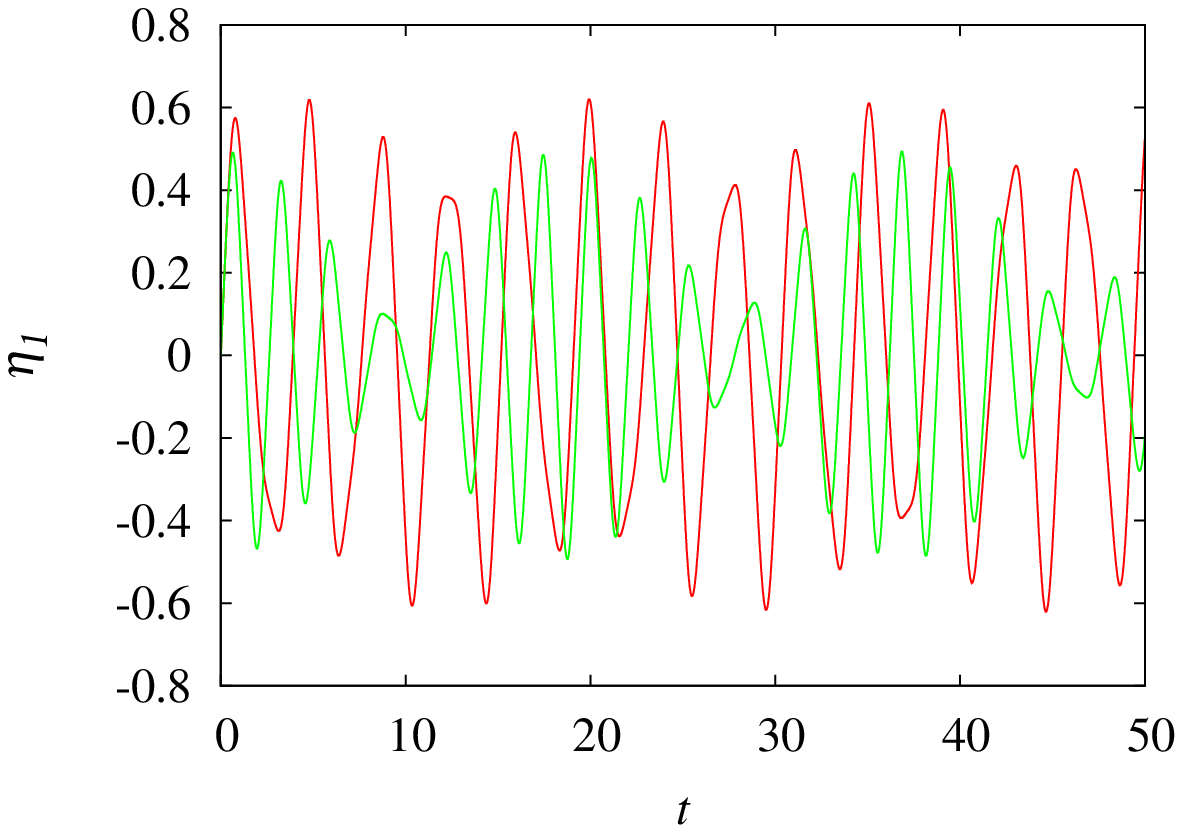}    \\
    \includegraphics[width=8.6cm,clip=true]{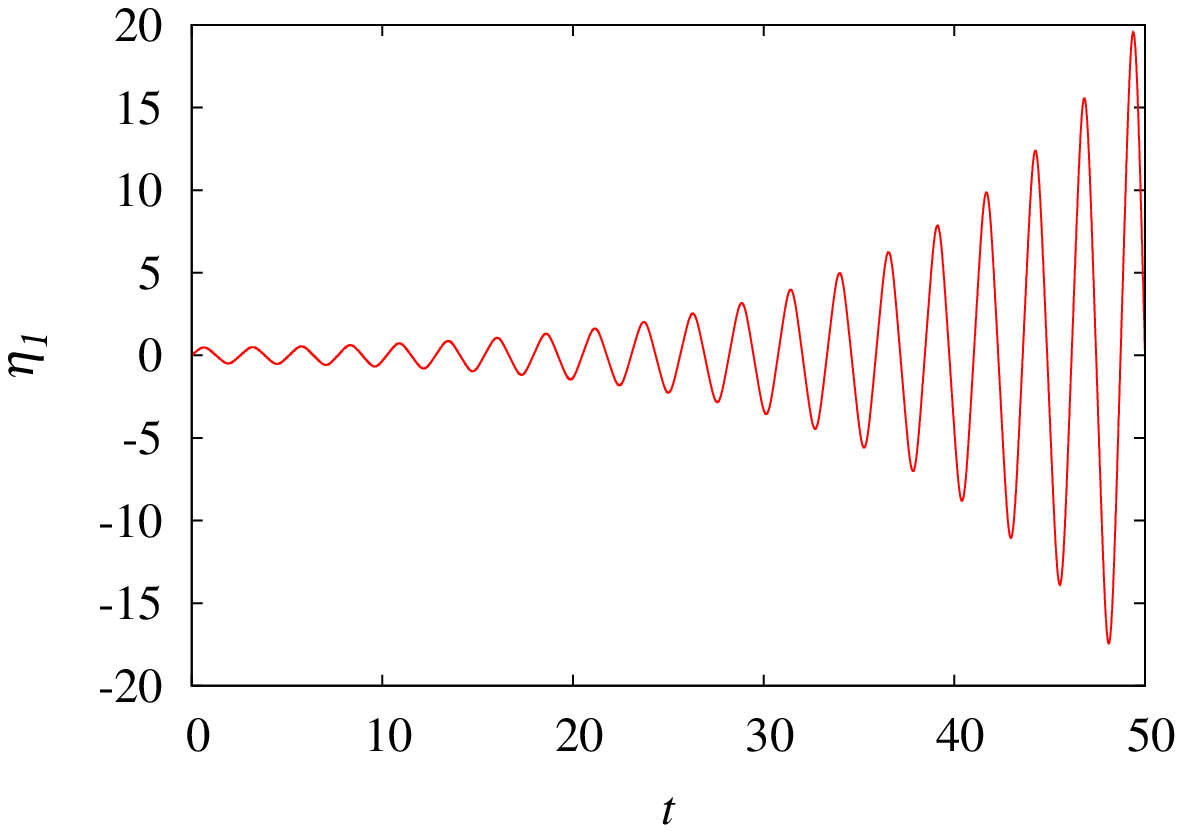}    
    \caption{Time dependence of the perturbations around the
      antisymmetric mode, $\chi_0$, for $N=2$: (Top) Perturbation of
      $\eta_1$ for $E/N=4$(red), 6.8(green). (Bottom) $\eta_1$ for
      $E/N=7.2$(red).  The perturbations develop beat behavior and
      become unstable for $E>7.12$. }
\label{fig:antiSym2}
\end{figure}
In the antisymmetric mode, $\eta_0=0$ and $\chi_0$ satisfies the
non-linear oscillator equation, \eqnn{antiSym}. Similarly to the
symmetric mode case, we need to consider the cases, $\omega^2(t)
=3\chi_0^2,3\chi_0^2+4$ for the equation \eqnn{hoLandau}.  Let us keep
the leading order term in $\chi_0$ expansion and deduce what happens:
Approximating $\chi_0$ by $\sqrt{E_1/2}\sin(2t)$, we find that
$\omega_0^2=3\chi_ 0^2+4 $ never satisfies the resonance condition,
\eqnn{condLandau}, while $\omega_0^2=3\chi_ 0^2$ satisfies it in the
region $3.4\simeq 2^8/(3\cdot 5^2)<E_1<2^8/3^3\simeq9.5$, so that there is
an unstable region.

While the above analytical argument used simple crude
approximations, they nevertheless give us insight as to the underlying
mechanism behind the stability and instability of the perturbations.
When the perturbation equations, \eqnn{fluct}, are integrated
numerically, we find that indeed the mode for $\omega^2=3\chi_ 0^2+4$
never becomes unstable, while that for $\omega^2=3\chi_ 0^2$ becomes
unstable for $E/N>7.12$. So the analytic argument recovers the rough
picture, but there is no upper bound to the energies for the
instability of the perturbations around the orbits, and this is
outside the region of the validity of the analytic approximation. The
transition from stable to unstable displays beat behavior just before
becoming unstable, which can be seen in \figno{antiSym2}.
\subsection{General periodic orbits for any $N$}
\label{sec:NStab} 
Let us now discuss the case for a general periodic orbit, for any $N$
and any boundary conditions. A periodic orbit $z_0$ satisfies the
equation, \eqnn{eomf}, for some $m_0$, as
\begin{equation}
    \label{eq:eomf0}
        \ddot  z_0 =     -\left(\omega^{(m_0)}\right)^2    z_0 -  z_0^3\quad.
\end{equation}
We expand around this basic solution as $q_j=q_{j,0}+q_{j,1}$, where
$q_{j,0}=\pm z_0$ or 0, due to \eqnn{cond}.
\begin{equation}
    \label{eq:fluctZQ}
         \ddot q_{j,1}= q_{j+1,1}+q_{j-1,1}-2q_{j,1}-3q_{j,0}^2q_{j,1}\quad.
\end{equation}
These equations are linear equations with respect to $q_{j,1}$, but
contain time dependent coefficients in $q_{j,0}$ and furthermore, $N$
$q_{j,1}$'s are coupled. When the amplitudes in the mode do {\em not}
contain zeros ({\it c.f.} \tabno{modesBC}), the equations can be further
simplified. In this case, using the normal mode coordinates for the
perturbations, $\mode {z_1}m$, we obtain the a form of the equations
which are decoupled,
\begin{equation}
    \label{eq:fluctZ}
     \mode{\ddot z_1}m =     -\left[\left(\mode\omega m\right)^2+3z_0^2
    \right]   \mode{z_1}m
    \quad.
\end{equation}
For a periodic solution of the $\phi^4$ lattice generated by $z_0$,
$N$ equations labeled by the normal mode
directions,  $m$, for the perturbations around the original periodic
solution.
The spectrum of the (harmonic) normal
modes $\mode\omega m$ enters the equation. We see that the $N=2$
symmetric and antisymmetric cases discussed above can be recognized as
special cases of these equations. It should be noted that the
unperturbed solution enters only as $z_0$ but this exists only for
certain values of $m_0$, as seen in \tabno{modesBC}. The equation
cleanly separates the role of the non-linear oscillatory mode in $z_0$
and the harmonic normal modes in the spectrum, $(\mode\omega 
m)^2$.
$N$ enters only through the spectrum. 

The above second order linear differential equation, \eqnn{fluctZ} ,
with a real periodic function as the coefficient is an example of Hill's
equation. Its two linearly independent solutions have one of the
properties below, given by {\em Floquet's theorem}\cite{HillEq}:
\begin{itemize}
  \item [(a)] The linearly independent solutions are of the form
    $e^{i\alpha t}p_+(t),e^{-i\alpha t}p_-(t)$, where, $p_\pm(t)$ are
    periodic functions of $t$ with the period $T$.
  \item [(b)] A non-trivial periodic solution, $p(t)$ exists.  Another
    solution, $f(t)$ has the property $f(t+T)=\pm
    f(t)+\theta p(t)$,  ($\theta$:  constant).
    % f(t)+\text{const.}\times p(t)$. 
%
 The period of $p(t)$ is $T$ or $2T$,
    and the sign in front of $f(t)$ is $+$ or $-$, respectively.
\end{itemize}
Here, $T$ is the (minimal) period of $z_0(t)^2$. Exponentially growing
perturbations exists if and only if the solutions are of type (a) with
a non-real $\alpha$. Solutions of type (a) with real $\alpha$ lead to
bounded perturbations. Solutions of type (b) lead to linearly growing
perturbations when $\theta\not=0$.

Let us discuss here the relation between the Lyapunov exponents,
perturbation equations, \eqnn{fluctZ}, and Floquet's theorem. Lyapunov
exponents measure the exponential rate at which the deviations from a
trajectory diverge from the solution. The trajectory needs not be
periodic. The number of Lyapunov exponents equals the dimension of the
phase space, $2N$, since the deviations can be made in any direction
in phase space. The behavior of perturbations around a periodic
solution may be obtained by solving the perturbation
equations. Intuitively, it should be expected that the growth rate of
the perturbations should be consistent with the Lyapunov exponents
averaged along the periodic solution. This shall be quantitatively
confirmed below. It should be noted that the computations involved in
obtaining the Lyapunov spectra and solving the perturbation equations
are quite different. To obtain the Lyapunov spectrum, in principle, we
need to solve the equations of motion and measure how different
solutions with close initial conditions diverge. In practice, in the
method we adopt, the equations of motion for the whole system are
solved while also tracking the evolution of vectors in the tangent
space, a procedure that requires $2N(N+1)$ coupled first order
equations to be solved. The whole spectrum of $2N$ exponents along
with the unperturbed periodic solution is
obtained this way, without using the perturbation equations,
\eqnn{fluctZ}. On the other hand, in perturbation theory, first, the
periodic solution is obtained and then, the $N$ perturbation equations
are solved, one by one.  Each equation should correspond to two
Lyapunov exponents. When the perturbations equation is of type (a) in
Floquet's theorem and $\alpha$ is not real, a perturbation can grow
exponentially and $\pm\Im\alpha$ should coincide with the Lyapunov
exponent pair $\pm\lambda$. In all other cases, the corresponding
Lyapunov exponents are zero. This will be explicitly seen below.  The
pairing property is consistent with the general property of Lyapunov
exponents. Interestingly, by independently solving one second order
differential equation for each perturbation mode, we recover the
Lyapunov exponents pair ($\pm\lambda$) by pair, including their
degeneracies.
To understand how the system works, we  illustrate this with a few concrete examples.
\subsubsection{General $N$, symmetric solution, periodic boundary
  conditions}
\label{sec:sym}
\begin{figure}[htbp]
    \centering
    \includegraphics[width=8.6cm,clip=true]{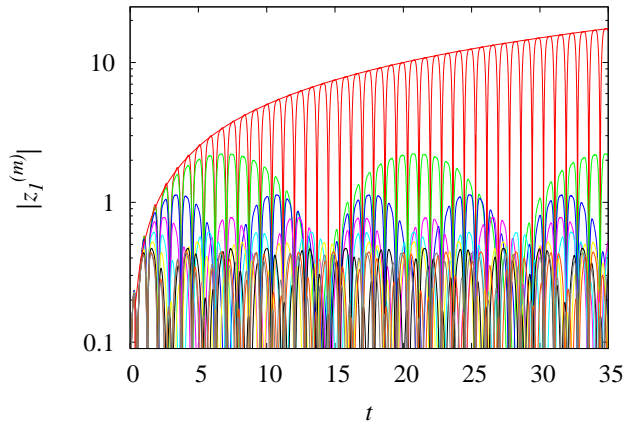}    
    \caption{Absolute values of the perturbations, $|\mode {z_1}m|,\
      (m=0,1,\cdots N/2)$ around the symmetric solution, $\mode
      {z_0}0$ as a function of time, $t$, for $N=16$, $E/N=100$.  All
      the different modes $m=0$ (red), 1(green), 2(blue), 3(magenta),
      4(cyan), 5(yellow), 6(black), 7(orange), $N/2=8$(grey) are
      shown. $t/2$ behavior is also shown (red), which matches well
      the linear, but {\em not} exponential, growth of $\mode {z_1}{
        0}$.  Individual modes, apart from $\mode{z_1}0$ are difficult
      to separate visually, but it can be seen that while $E/N$ is
      relatively large, none of the modes have exponentially
      increasing amplitudes. At this $E/N$, the modes have decreasing
      amplitudes in the order, $m=0,1,2,\cdots N/2$.}
    \label{fig:stabN}
\end{figure}
When the boundary conditions are periodic, the symmetric solution,
where all coordinates and momenta move in unison, $q_j=q_k,p_j=p_k$
(any $j,k$) is a solution for general $N$. To analyze the
perturbations to the periodic orbit through the equations,
\eqnn{fluctZ}, we need the spectrum of the normal modes for the
harmonic theory, which,   for periodic boundary conditions is
\begin{equation}
    \label{eq:spectrumPBC}
    \mode{\omega}m=2\sin\pi{m\over N},\quad
    m=0,1,\cdots N-1.
\end{equation}
This spectrum is doubly degenerate except for $m=0$ and $N/2$, the
latter only when $N$ is even.  So there are only $N/2+1$ or $(N+1)/2$
independent perturbation equations, \eqnn{fluctZ}, depending on
whether $N$ is even or odd.  When the equations \eqnn{fluctZ} are
integrated, we find no instabilities for small or large $E/N$ for any
$N$.  

An example is shown in \figno{stabN} for $N=16$ lattice at $E/N=100$.
While the energy of the system is relatively large, none of the
perturbations grows exponentially.  This result is quite consistent
with the Lyapunov exponents being immeasurably small numerically, for
any $N$ and $E/N$\cite{HA1}.
In this case, the perturbations for the modes $m=1,2,\ldots,N/2$
belong to case (a) of Floquet's theorem, with real values of
$\alpha$. The distinct feature of perturbations with two periods can
be observed in the figure.  The mode $m=0$ is of case (b) of
Floquet's theorem with $\theta\neq0$. It should be noted that $m=0$
coincides with the linearized normal mode of the original periodic
solution, which exists for any energy. The perturbation in this
direction does not grow exponentially so that the corresponding
Lyapunov exponents are zero, yet grows linearly for the following
reason. If $\theta=0$, we would have two perturbations which are
periodic when shifted with the period of the unperturbed solution
$z_0$. This would lead to a slightly perturbed periodic solution with
the same period. However, non-linear periodic solutions change both
the period and the trajectory shape with the energy, which leads to a
linear growth with respect to $t$ in the perturbation. Clearly, this
argument is not restricted to the symmetric periodic orbit and we
shall see that this property holds for all the examples we study.
\subsubsection{General even $N$, antisymmetric solution, periodic boundary
  conditions}
\begin{figure}[htbp]
    \centering
    \includegraphics[width=8.6cm,clip=true]{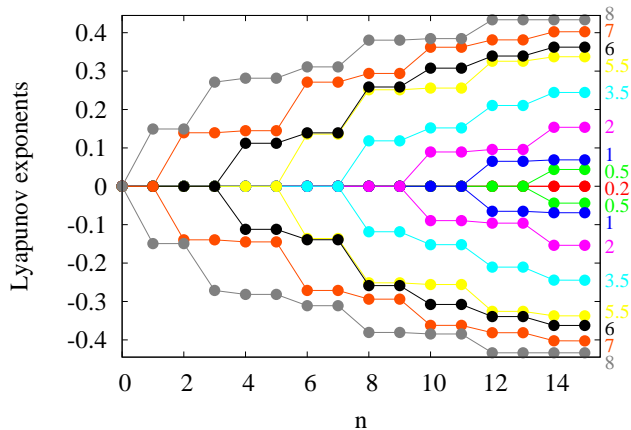}    
    \caption{Lyapunov exponents $\{\lambda_n\}$ of $N=16$ antisymmetric
    orbits, for $E/N=0.2$(red), 0.5(green), 1(blue), 2(magenta),
    3.5(cyan), 5.5(yellow), 6(black), 7(orange), 8(grey) and labeled
    on the right hand side.  The
    Lyapunov exponents are plotted in the increasing order of their
    magnitudes. 
    The whole spectrum is always invariant under the reflection
    $\lambda_n\leftrightarrow-\lambda_n$ , as it should be.  More and
    more Lyapunov exponents are seen to become non-zero, corresponding
    to more perturbation modes around the periodic orbit becoming
    unstable.  The exponents are seen to become nonzero in pairs (of
    $\pm \lambda _n$ pairs) with the same values, except for
    $\lambda_0=0$ and the last exponent to become non-zero
    $\pm\lambda_3$ for $E/N=8$. The exponents, including their
    degeneracy, reflect the properties of the perturbations.  }
    \label{fig:lyapA}
\end{figure}
\begin{figure}[htbp]
    \centering
    \includegraphics[width=8.6cm,clip=true]{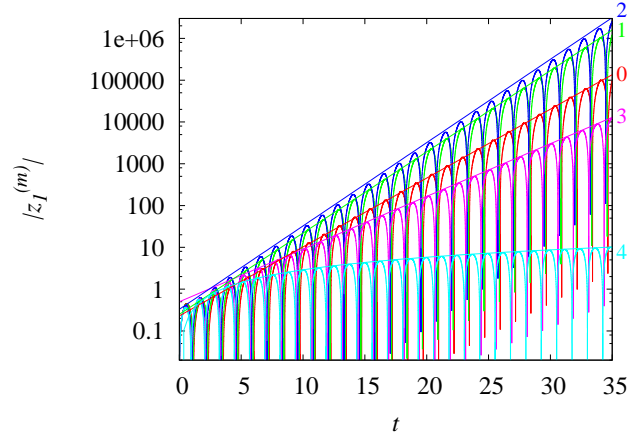}    
    \caption{Perturbation about the antisymmetric orbit for $N=8$,
      $E/N=9$: Modes for $m=0$(red),1(green), 2(blue), 3(magenta),
      4(cyan) are shown (also labeled on the right hand side), which
      grow exponentially with time except for mode $m=4$, which grows
      linearly.  $\exp(\mode\lambda p t)$ are also shown (in color
      corresponding to the mode) and agree excellently with the growth
      of the perturbations for $m=0,1,2,3$.  The corresponding
      strictly positive exponents are $\mode\lambda0=0.3795,
      \mode\lambda1=0.4476, \mode\lambda2=0.4591,\mode\lambda3=0.2903$
      and all exponents are doubly degenerate in the Lyapunov
      spectrum, except for $\mode\lambda0$. $0.29t$ is shown and
      agrees well with the linear growth of the perturbation for $m=4$. }
    \label{fig:expGrowth}
\end{figure}

The situation is much more interesting for the antisymmetric periodic
solution, $q_j=-q_{j+1}, p_j=-p_{j+1}$ for general even $N$. When
$N=2$, it was seen that for large enough $E/N$, the orbit becomes
unstable and one (of the two) fluctuation equations had a parametric
resonance. For general $N$, there are $N/2+1$ inequivalent
perturbation modes, as explained above. As we increase $E/N$, the
perturbation modes become unstable, one inequivalent mode by one. This
can be seen as the solutions to the perturbation equations,
\eqnn{fluctZ}, developing exponentially growing behavior, as in
\figno{expGrowth}, which is also evidenced in the Lyapunov spectrum,
\figno{lyapA}. The number of non-zero exponents can be seen to
increase systematically as $E/N$ is increased.

The equations for the perturbations, \eqnn{fluctZ}, also reflect the
degeneracy of the spectrum, \eqnn{spectrumPBC}. From the computations
of Lyapunov spectra, this degeneracy is a priori not obvious, but it
is indeed reflected in the Lyapunov spectra, as seen in
\figno{lyapA}. The modes become non-zero in pairs, corresponding to
identical Lyapunov exponents (and $(-1)$ times them), except for two
modes, as we now explain.  The two {\em non}-degenerate modes
$m=0,N/2$ correspond to the symmetric and antisymmetric periodic
orbits respectively. The unperturbed orbits $z_0$ satisfy the
non-linear oscillator equation \eqnn{eomf0} with $(\mode\omega 0)^2=0$
or $(\mode\omega {N/2})^2=4$, neither of which depends on $N$.  So the
equations for these two perturbation modes and hence their behavior
are independent of $N$.  The mode for $m=0$ corresponds to a
non-degenerate non-zero Lyapunov exponent pair, whose value is
independent of $N$. This was found in \cite{HA1} and the reason for
this is now clear.  This mode becomes unstable at the highest energy,
amongst the modes.  
$m=N/2$ mode direction coincides with the periodic orbit which we
perturbed around, and the perturbation grows linearly for the reason
explained at the end of \sect{sym}.  In regards to Floquet's theorem,
perturbation equation for $m=N/2$ mode is of type (b) with
$\theta\neq0$ and other modes are of type (a). 
In general, there needs to be a mode associated with the pair of zero
Lyapunov exponents for any trajectory, including periodic orbits, in
particular, not just the antisymmetric orbit. The perturbation of
along the original periodic solution itself performs this role, and
it is evident here  that it is the only mode that can, in general.
The spectrum of Lyapunov exponents should correspond to the rate of
exponential growth of the perturbations with time, as discussed above.
The spectrum computed independently is seen in \figno{expGrowth} to
agree with the growth rate of perturbations quantitatively. 
% The Lyapunov spectrum is computed by solving the
% equations of motion for the whole system and also tracking the
% evolution of vectors in the tangent space, a procedure that requires
% $2N(N+1)$ first order equations to be solved. This evidently agrees
% with the spectrum computed independently, as can be seen in
% \figno{expGrowth}.

\begin{figure}[htbp]
    \centering
    \includegraphics[width=8.6cm,clip=true]{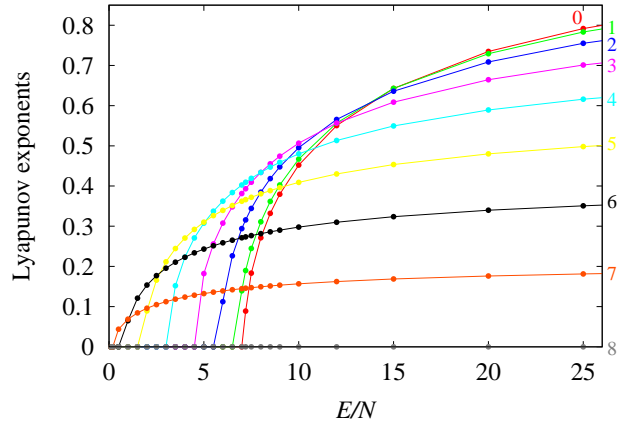}    
    \caption{The dependence on $E/N$ of the Lyapunov exponents along
      the antisymmetric orbit for $N=16$, corresponding to the modes
      $m=0$ (red),1(green),2(blue), 3(yellow), 4(cyan), 5(magenta),
      6(black), 7(orange), and $N/2$=8 (grey) (la belled in the
      plot). Only positive Lyapunov exponents ($\lambda\geq0$) are
      shown, since the negative exponents are identical, except for
      the sign.}
    \label{fig:lyapE}
\end{figure}

For perturbations around symmetric orbit, at $E/N=100, N=16$, the size
of the amplitudes corresponding to the modes were in descending order
with respect to the modes $m=0,1,\cdots,N/2$, as seen in
\figno{stabN}.  However, in the example of the antisymmetric
$E/N=9,N=8$, no such simple ordering exists. It is interesting to see
how the growth of perturbations, as characterized by Lyapunov
exponents, depend on the mode for a given $E/N$. 
This is shown for the antisymmetric orbit for $N=16$ in \figno{lyapE},
where it is seen that for large $E/N$, the ordering is similar to what
was seen for the symmetric orbit. For perturbations around the
antisymmetric orbit, the modes $m=N/2-1,N/2-2,\cdots,2,1,0$ become
unstable one by one as we increase $E/N$. However, as we increase the
energy, eventually, the size of the Lyapunov exponent is in the
reverse order they became non-zero. This property holds for all $N$ we
have investigated, and a mathematical structure presumably exists
behind it, which still needs to be investigated.
\section{Extension of the construction and stability analysis to models
   with different on-site potentials}
\label{sec:extension}
As noted in \sect{construction}, the construction of the periodic
orbits can be applied to general one dimensional lattice theories,
provided the inter-site couplings are harmonic. The theory can be more
general in two ways, the harmonic part of the Hamiltonian can be
different, and also, the on-site potential can be different.  If the
normal nodes for the harmonic theory without the on-site potential, is
known, we can use the condition \eqnn{cond} to systematically find
periodic solutions.  We now consider Hamiltonians without changing the
inter-site potential, but with different on-site potentials,
$\Phi(q)$.  The Hamiltonian for these theories are,
\begin{equation}
    \label{eq:hamG} H=\sum_{j=1}^N
   {p_j^2\over2}+\sum_{j=1}^{N
      -1}{(q_{j+1}-q_j)^2\over2}+H_{\rm B} +\sum_{j=1}^N \Phi(q_j)
    \quad,
\end{equation}
The equations of motion for the theory are coupled $2N$ first order
non-linear differential equations, in general.  When a normal mode can
be extended to the non-linear theory, the same construction yields the
equations of motion, which is a single second order differential
equation.
\begin{equation}
    \label{eq:generalEOM}
    \ddot  z =     -(\omega^{(p)})^2    z - {d\Phi\over dz}(z)\quad,
\end{equation}
The condition for such a reduction to apply is identical to
\eqnn{cond}, as long as the potential is even with respect to $z$,
$\Phi(-z) =\Phi(z)$. When the potential is not even, only the
symmetric mode, with all coordinates having the same value, is
allowed. This mode might be incompatible with the boundary conditions
of the theory. Strictly speaking, when the potential is not even, a
mode with all the non-zero amplitudes having the same value is
allowed, which is more general, in principle, than the symmetric mode.
However, only the symmetric mode resides in this category for the
Hamiltonians considered here (see, \tabno{modes}). When the trajectory
is bounded, it is periodic, even though the reduction of the equations
of motion is applicable even when the motion is unbounded.  The
potential does not need to be a monomial, or even a polynomial. If the
inter-site potentials are changed, the normal modes for the harmonic
part will be different, but the same principle applies to extending
the normal modes to the non-linear theory. We parenthetically point
out one exception to these considerations, the case $\Phi(z)={\rm
  const.}\times z^2$. In this case, all the normal modes, with any
amplitude, extend to the theory with this on-site potential, but the
equations of motion are linear.

The perturbations around a solution, $(q_{j,0})$, can be analyzed
analogously to the $\phi^4$ theory.  The perturbations, $(q_{j,1})$,
satisfy
\begin{equation}
    \label{eq:fluctZQG}
         \ddot q_{j,1}= q_{j+1,1}+q_{j-1,1}-2q_{j,1}-{d^2\over dz^2}\Phi(q_{j,0})q_{j,1}\quad.
\end{equation}
These equations are are applicable in general, given any periodic
solution, $(q_{j,0})$, obtained using \eqnn{generalEOM}.  They can be
further simplified to the following equations when $\Phi(q_{j,0})$ is
independent of $j$, which occurs when no zeros exist in the amplitudes
of the modes, as in the $\phi^4$ theory.
\begin{equation}
    \label{eq:fluctZG}
    \mode{\ddot z_1}m =     -\left[\left(\mode\omega
        m\right)^2+{d^2\over dz^2}\Phi(z_0)    \right]   \mode{z_1}m
    \quad,\quad
    m=0,1,2,\cdots N-1.
\end{equation}
If the harmonic part of the Hamiltonian is the same as those of the
coupled oscillator \eqnn{ham}, for periodic, fixed or free boundary
conditions, the general solutions to the conditions, \eqnn{cond}, have
been analyzed in \sect{construction}. 

We now examine some concrete examples:
\paragraph{Cubic and quartic potential}
\begin{figure}[htbp]
    \centering
    \includegraphics[width=8.6cm,clip=true]{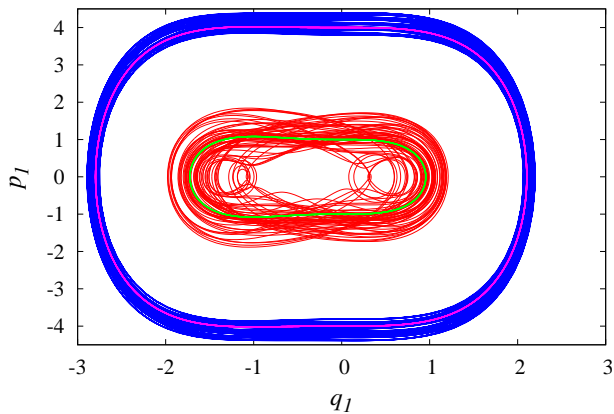}        
    \caption{Periodic orbits in phase space for the potential
      $\Phi(q)=q^3/3+q^4/4$ and their perturbations, for $N=4$: The
      symmetric trajectory $(q_1,p_1)$ for $E/N=0.5$(green) , 8
      (magenta) and their perturbations (red, blue, respectively).
      Evidently, $E/N=0.5$ orbit is unstable, while $E/N=8$ orbit is
      not, which is agrees with the Lyapunov spectra computed along
      these orbits. The orbits are not symmetric with respect to
      $q_1\leftrightarrow-q_1$ reflection and the orbits for different
      energies are seen to be quite dissimilar in shape.  Also, the
      orbits both differ from the harmonic oscillator orbit, an
      ellipse.  All trajectories were started with $q_j=0 \ (\hbox{any
      } j)$ and were followed for the same amount of time, $\Delta
      t=400$. Perturbed orbits were obtained by increasing the initial
      $p_1$ values by 10\,\%. }
    \label{fig:phi34Orbit}
\end{figure}
Let us consider the example of a potential, by adding a cubic term to
the quartic potential of $\phi^4$ theory,
\begin{equation}
    \label{eq:phi34}
    \Phi(q)={q^3\over3}+{q^4\over4}\quad.
\end{equation}
This potential is different from the $\phi^4$ theory potential in that
it is not a monomial, and further does not have the reflection
symmetry $q\leftrightarrow -q$. Therefore, only the symmetric mode, as
explained above, can be extended to the non-linear theory, which is a
solution for any $N$, when the boundary conditions are periodic or
free. Some periodic orbits for $N=4$ lattice with periodic boundary
conditions are shown in \figno{phi34Orbit} at $E/N=0.5,8$ along with
their perturbed trajectories. The former is unstable and the latter is
stable. There are interesting qualitative differences in the behavior,
when compared to the quartic potential trajectories analyzed in the
previous section. In that case, the symmetric orbit was always stable.
Furthermore, when the orbit became unstable, increasing $E/N$ only
made it less stable (see \figno{lyapE}). With the current potential,
the symmetric orbit becomes unstable at lower energies and becomes
stable at higher energies.  This behavior might seem intuitively
strange at first sight, especially since increasing the energy might
naively seem to enhance the instability of the orbits. However, at
higher energies, the orbit samples potentials at higher energies on
average, which is governed by the quartic behavior. Therefore, the
essential differences from the quartic potential case become more
pronounced at lower energies.  As in the previous section, the the
rates of growth of the perturbation modes quantitatively agree with
the Lyapunov spectrum, which has been computed independently.
\paragraph{Trigonometric potential}
\begin{figure}[htbp]
    \centering
    \includegraphics[width=8.6cm,clip=true]{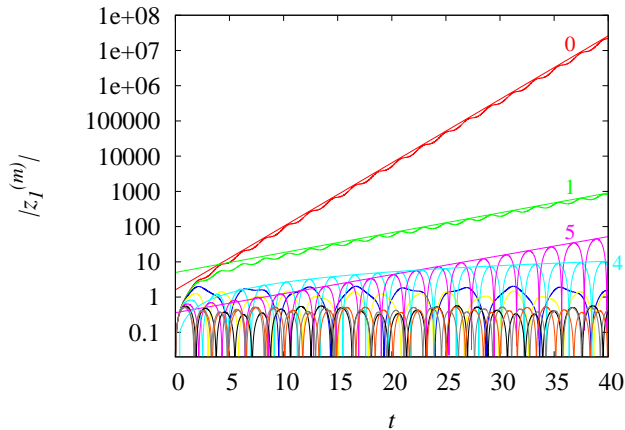}        
    \caption{Trigonometric potential: Absolute value of the
      perturbations, $|\mode {z_1}m|,\ (m=0,1,\cdots N/2)$ around the
      mode 2a periodic solution, $ {z_0}$, with $\omega_0^2=2$,  as a function
      of time, $t$, for $N=16$, $E/N=10$.  All the different modes
      $m=0$ (red),1(green),2(blue), 3(yellow), 4(cyan), 5(magenta),
      6(black), 7(orange), N/2=8 (grey) are shown. Modes $m=0,1,5$
      have exponential growing perturbations. Using the Lyapunov
      spectrum computed independently, $\exp(\mode\lambda m t)$
      corresponding to these three modes are shown, and the growth
      rate and the Lyapunov exponents are in excellent agreement. The
      corresponding exponents are $\mode\lambda0= 0.4150,
      \mode\lambda1= 0.1230, \mode\lambda5= 0.1243$ and
      $\mode\lambda{1,5}$ are doubly degenerate in the Lyapunov
      spectrum.  Perturbation for mode $m=4$ grows linearly and its
      growth agrees well with $0.34t$, which is also shown.}
    \label{fig:lyapCos}
\end{figure}
As a final example, let us analyze 
\begin{equation}
    \label{eq:cosPotential}
    \Phi(q)=1-\cos(q)\quad.
\end{equation}
This case is qualitatively different from the previous examples; the
potential is {\it not} a polynomial function and the potential is
bounded, so that there exist unbounded orbits. For this case, we
analyze the perturbations around the mode based on mode 2a in
\tabno{modes}, for the $N=16$ lattice with periodic boundary
conditions (\tabno{modesBC}). This mode is different from the
symmetric and the antisymmetric mode, but can be analyzed in the same
fashion, using the theoretical structure introduced above.

In \figno{lyapCos}, the time dependence of the absolute values of all
the perturbation modes for $E/N=10$ are shown. At this energy, this
mode is bounded and periodic due to the quadratic part of the
potential in the Hamiltonian, \eqnn{hamG}. The symmetric mode would 
lead to {\em un}bounded trajectories at the same energy.  There are three
unstable modes, two of which run away and will not oscillate around
zero, and one unstable mode that does not run away. Such run away
modes can exist when $\left(\mode\omega m\right)^2<1$ so that the time
dependent frequencies in \eqnn{fluctZG} can become imaginary, which
corresponds to $m/N<1/6$. In this example, $m=0,1$ modes run away and
$m=5$ mode has a parametric resonance type instability.
These perturbations are of type (a) in Floquet's theorem with $\alpha$
non-real. The run away solutions correspond asymptotically to periodic
solutions oscillating around a non-zero value with exponentially
growing amplitudes. The deviations from the simple exponential growth
behavior for small $t$ is due to the contribution of the exponentially
decaying solutions.  $m=4$ mode corresponds to the mode 2a of the
unperturbed periodic solution and grows linearly, as can be seen in
\figno{lyapCos}.  This agrees with the general argument given at the
end of \sect{sym}, and the solution is of type (b) in Floquet's theorem with
$\theta\neq0$.
The Lyapunov spectrum, computed independently, confirms that there are
five strictly positive exponents and their values can be seen to agree
quite well with the growth of the perturbations, as can be seen in
\figno{lyapCos}. Lyapunov exponents for the $m=1,5$ modes are doubly
degenerate, as explained in the previous section.
\section{Discussion}
\label{sec:disc}
In this work, we systematically constructed periodic orbits of the
$\phi^4$ theory by extending the normal modes of the harmonic limit of
the model.
 The stability of the periodic orbits were analyzed, quantitatively
relating the Lyapunov exponents to each modes.  Properties of the
Lyapunov spectrum, such as the degeneracy of the exponents and their
relation to the harmonic spectrum were clarified. 
While some of the general properties have been known and explicit
results have been derived for some other
models\cite{NLNMRev,ChechinRev}, these questions have not been studied
for the $\phi^4$ theory. We believe that the results complement the
previous results in other models such as the FPU model in an
interesting way, considering their different dynamical behaviors.
Also, importantly, the $\phi^4$ theory is a prototypical model in this
class in various fields of physics.  Furthermore, by showing how the
various aspects come together explicitly, our results can hopefully
serve as a concrete basis for future research.
The systematic construction of the periodic orbits and their stability
analysis are applicable to other models with harmonic inter-site and
non-linear on-site potentials, and we studied how this can be done
with models having qualitatively different behavior from the $\phi^4$
theory.
%
% Through the process, we have shown how the normal modes of harmonic
% models can be extended to non-linear models with on-site potentials,
% or perhaps more importantly, when and why they can not be extended.
It should be noted that for this class of models, these periodic orbits
exhaust the solutions in which all the coordinates move in
synchronization and the overall amplitude is arbitrary, by
construction.
%   Using the methods
% developed here, we investigated some periodic orbits in the $\phi^4$
% theory, in detail. We further analyzed their stability, by
% systematically setting up perturbation theory around these orbits and
% clarifying the relation of the perturbation modes to the Lyapunov
% exponents quantitatively. Furthermore, 
% We also analyzed periodic orbits with other on-site potentials, which
% had qualitatively different properties from the quartic potential of
% the $\phi^4$ theory.  
We found a fascinating consistent picture that
ties together the physics of the periodic orbits, perturbation around
them and the Lyapunov spectra.  One can obtain Lyapunov exponents
around these orbits, one $\pm\lambda$ pair by pair, by solving a
single second order differential equation at a time, rather than
solving for the whole system, providing a clear picture of the system.

There are several directions to be further investigated. The $\phi^4$
theory can be studied more deeply, using extension of normal modes and
boundary conditions, other than those studied here. Also, given the
general construction presented here, the dynamics of periodic orbits
in models with different on-site potentials and their stability can be
studied.  While we have studied synchronous periodic orbits with
arbitrary amplitudes for these class of theories, more general
periodic solutions exist, sometimes referred to as higher dimensional
bushes\cite{bush0,bush1}. 
%  a question remains as to what other kinds of periodic orbits exist.  
% One might also wonder whether there also exist synchronous orbits for a particular
% amplitude or above a certain amplitude, for instance.  Furthermore, in
% the harmonic theory, we can add normal modes with different
% frequencies to construct another periodic solution, which is {\em not}
% a normal mode. Non-linear theories prohibit such a simple
% construction, but it would be interesting to clarify if such solutions
% can be extended to the non-linear theory and if so, under which
% conditions. 
Behavior of periodic solutions and their dependence on the properties
of the on-site potential raises intriguing questions.  The link
between the stability and instability of the periodic orbits and the
behavior of Lyapunov exponents can be studied at a deeper level.
%  For
% instance, there needs to exist at least one stable perturbation around
% any periodic orbit, on general grounds.  However, this is not obvious
% from the individual perturbation equations, \eqnn{fluctZG}, and there
% are properties that can only be understood from viewing the whole set
% of equations.  Such 
Consistency requirements and the behavior of Lyapunov exponents, as
seen in \figno{lyapE}, hint at a beautiful underlying mathematical
structure, which is intriguing.
We believe that the interesting
subject matter studied here brings together various fields in
classical dynamics; analytic aspects, such as parametric resonance,
geometric aspects such as the periodic trajectories and the Lyapunov
spectrum, and applied physics.  
We hope that the concrete results presented here for the $\phi^4$
lattice theory and other models leads to further progress in the
field. 

\acknowledgments We would like to
thank William Hoover for the collaboration that started this project,
constructive comments, and encouragement.  K.A. was supported in part
by the Grant--in--Aid for Scientific Research (\#15K05217) 
from the Japan Society for the Promotion of Science (JSPS),
and a grant from Keio University.
\end{document}